\documentclass[aps,twocolumn,prx,preprintnumbers,showpacs,amsmath,amssymb,superscriptaddress,longbibliography]{revtex4-1}
\usepackage[pdftex, colorlinks, citecolor=blue]{hyperref}   
\usepackage{graphicx}
\usepackage{dcolumn}
\usepackage{threeparttable}
\usepackage{multirow}
\usepackage{booktabs}
\usepackage{txfonts}
\usepackage{xcolor}
\usepackage{bm}
\usepackage{amssymb}
\usepackage{amsmath}
\usepackage{latexsym}
\usepackage{epsfig}
\usepackage{amsbsy}
\usepackage{array}
\usepackage{tabularx}
\usepackage{esvect} 
\usepackage{extarrows}
\usepackage{float}
\usepackage[british]{babel}

\begin{document}

\title{Origin of the charge density wave state in BaFe$_2$Al$_9$}

\author{Yuping Li}
\thanks{These authors contribute equally to this work.}
\affiliation{Key Laboratory for Anisotropy and Texture of Materials (Ministry of Education),
		School of Materials Science and Engineering, Northeastern University, Shenyang 110819, China}
\affiliation{%
Shenyang National Laboratory for Materials Science, Institute of Metal Research, Chinese Academy of Sciences, 110016 Shenyang, China
}%

\author{Mingfeng Liu}
\thanks{These authors contribute equally to this work.}
\affiliation{%
Shenyang National Laboratory for Materials Science, Institute of Metal Research, Chinese Academy of Sciences, 110016 Shenyang, China
}%

\author{Jiangxu Li}
\email{jxli15s@imr.ac.cn}
\affiliation{%
Shenyang National Laboratory for Materials Science, Institute of Metal Research, Chinese Academy of Sciences, 110016 Shenyang, China
}%

\author{Jiantao Wang}
\affiliation{%
Shenyang National Laboratory for Materials Science, Institute of Metal Research, Chinese Academy of Sciences, 110016 Shenyang, China
}%
\affiliation{%
School of Materials Science and Engineering, University of Science and Technology of China, 110016 Shenyang, China
}%

\author{Junwen Lai}
\affiliation{%
Shenyang National Laboratory for Materials Science, Institute of Metal Research, Chinese Academy of Sciences, 110016 Shenyang, China
}%
\affiliation{%
School of Materials Science and Engineering, University of Science and Technology of China, 110016 Shenyang, China
}%

\author{Dongchang He}
\affiliation{%
Shenyang National Laboratory for Materials Science, Institute of Metal Research, Chinese Academy of Sciences, 110016 Shenyang, China
}%
\affiliation{%
School of Materials Science and Engineering, University of Science and Technology of China, 110016 Shenyang, China
}%

\author{Ruizhi Qiu}
\affiliation{%
Institute of Materials, China Academy of Engineering Physics, Mianyang 621907, China
}%

\author{Yan Sun}
\affiliation{%
Shenyang National Laboratory for Materials Science, Institute of Metal Research, Chinese Academy of Sciences, 110016 Shenyang, China
}%

\author{Xing-Qiu Chen}%
\email{xingqiu.chen@imr.ac.cn}
\affiliation{%
Shenyang National Laboratory for Materials Science, Institute of Metal Research, Chinese Academy of Sciences, 110016 Shenyang, China
}%

\author{Peitao Liu}%
\email{ptliu@imr.ac.cn}
\affiliation{%
Shenyang National Laboratory for Materials Science, Institute of Metal Research, Chinese Academy of Sciences, 110016 Shenyang, China
}%

\begin{abstract}
Recently, a first-order phase transition associated with charge density wave (CDW)
has been observed at low temperatures in intermetallic compound BaFe$_2$Al$_9$.
However, this transition is absent in its isostructural sister compound  BaCo$_2$Al$_9$.
Consequently, an intriguing question arises as to the underlying factors
that differentiate BaFe$_2$Al$_9$ from BaCo$_2$Al$_9$ and drive the CDW transition in BaFe$_2$Al$_9$.
Here, we set out to address this question by conducting a comparative
\emph{ab initio} study of the electronic structures, lattice dynamics, \textcolor{black}{and electron-phonon interactions} of their high-temperature phases.
We find that both compounds are dynamically stable with similar phonon dispersions.
The electronic structure calculations reveal that both compounds are nonmagnetic metals; however,
they exhibit distinct band structures around the Fermi level.
In particular, BaFe$_2$Al$_9$ exhibits a higher density of states at the Fermi level with dominant partially filled Fe-$3d$ states
and a more intricate Fermi surface.
This leads to an electronic instability of BaFe$_2$Al$_9$ toward the CDW transition,
which is manifested by the diverged electronic susceptibility at the CDW wave vector $\mathbf{q}_{\rm CDW}=(0.5, 0, 0.3)$,
observable in both the real and imaginary parts.
Conversely, BaCo$_2$Al$_9$ does not display such behavior, aligning well with experimental observations.
\textcolor{black}{
Although the electron-phonon interactions in BaFe$_2$Al$_9$ surpass those in BaCo$_2$Al$_9$ by two orders of magnitude,
the strength is relatively weak at the CDW wave vector, suggesting that the CDW in BaFe$_2$Al$_9$ is primarily driven by electronic factors.}
\end{abstract}

\maketitle

\section{Introduction}

The charge density wave (CDW) is a structural phase transition
involving a periodic modulation of both the electron density and ionic positions
below a critical temperature~\cite{RevModPhys.60.1129,Zong2021}.
It is related to many intriguing physical properties
including metal-to-insulator transitions, nonlinear electrical transports,
unconventional superconductors, and intertwined electronic
orders~\cite{PhysRevB.35.6348,Zybtsev2010, RevModPhys.87.457,PhysRevB.87.115135,Chen_2016,Teng2022,Zheng2022}.
The simplest CDW is the Peierls instability~\cite{10.1093/acprof:oso/9780198507819.001.0001},
which appears in one-dimensional (1D) system due to the Fermi surface nesting (FSN).
However, the FSN mechanism was shown to rarely take a decisive role in real materials,
even for quasi-1D systems, since the divergence in the real part of the electronic susceptibility
is fragile against effects such as temperature, imperfect nesting, or scattering~\cite{PhysRevB.77.165135,Rossnagel_2011,2015Classification,Zhu2017}.
The CDW can also manifest in systems with higher dimensions,
wherein the CDW formation mechanisms tend to be more intricate and typically specific to the materials
involved~\cite{Rossnagel_2011,2015Classification,Zhu2017}.
These materials include two-dimensional (2D) layered transition-metal dichalcogenides such as NbSe$_{2}$, TiSe$_{2}$, TaS$_{2}$,
VSe$_{2}$, and VTe$_{2}$~\cite{PhysRevB.94.045131,PhysRevB.98.045114,PhysRevB.101.235405,
PhysRevResearch.5.013218,PhysRevLett.131.196401,PhysRevLett.121.196402,LASEK2021100523},
quasi-2D kagome materials AV$_3$Sb$_5$ (A=K, Rb, Cs)~\cite{10.1093/nsr/nwac199,PhysRevX.11.031050, PhysRevLett.127.236401,PhysRevLett.132.096101,PhysRevLett.127.046401},
kagome magnet FeGe~\cite{Miao_2023,Wanxiangang},
quasi-2D rare-earth tritelluride family~\cite{Zong2021},
and
5$f$-electron uranium~\cite{U_review_1994,Qiu_2016,Xie_2022}
as well as
kagome intermetallic ScV$_6$Sn$_6$~\cite{PhysRevLett.129.216402,PhysRevLett.130.266402,2023NatCo} with three-dimensional (3D) character.

Recently, an unusual occurrence of a CDW was experimentally reported
in a ternary aluminide compound BaFe$_2$Al$_9$ with 3D structural  networks~\cite{2021_BaFeAl}.
Different from the typical CDW phase transition, the CDW transition in BaFe$_2$Al$_9$ was found to be of first order,
as evidenced by a sudden change in magnetic susceptibility, electrical resistivity, and lattice parameters at a critical temperature of approximately 100 K~\cite{2021_BaFeAl}.
The structural modulation associated with the CDW has been resolved from the refinement of the neutron diffraction data
revealing a dramatic increase in Fe and Ba atomic displacement parameters,
and the low-temperature single-crystal x-ray diffraction showing additional superlattice peaks at a wave vector (0.5, 0, $q_z$) ($q_z$$\simeq$0.302)~\cite{2021_BaFeAl}.
An additional intriguing aspect of BaFe$_2$Al$_9$ is the absence of magnetic order at low temperature~\cite{2021_BaFeAl}.
This is somewhat unusual given that the  Fe-$3d$ orbitals are partially filled.
It is worth noting that BaCo$_2$Al$_9$ is isostructural to BaFe$_2$Al$_9$, but interestingly, no CDW phase transition was detected in BaCo$_2$Al$_9$~\cite{2021_BaFeAl}.
This naturally leads to the question of what sets BaFe$_2$Al$_9$ apart from BaCo$_2$Al$_9$
and what is the underlying cause of the CDW transition in BaFe$_2$Al$_9$.

Motivated by the different characteristics of the two isostructural compounds,
in this work we conducted a comparative
\emph{ab initio} study of the electronic structures, lattice dynamics, \textcolor{black}{and electron-phonon couplings (EPC)} of their high-temperature phases.
We find that both compounds exhibit similar phonon dispersions without the presence of imaginary phonon modes.
However, the two compounds exhibit distinct electronic band structures around the Fermi level.
This is due to the fact that the Fe-$3d$ orbitals are partially filled in BaFe$_2$Al$_9$,
whereas in BaCo$_2$Al$_9$ the Co-$3d$ orbitals are fully occupied.
This leads BaFe$_2$Al$_9$ to exhibit a relatively high density of states at the Fermi level and a more complex Fermi surface.
The electronic susceptibility calculations reveal the divergence in both the real and imaginary parts
at the CDW wave vector $(0.5, 0, 0.3)$ in BaFe$_2$Al$_9$,
which is, however, absent in BaCo$_2$Al$_9$.
\textcolor{black}{Additionally, at the CDW wave vector a weak EPC was obtained.}
Our calculations agree well with the experimental observations and
provide compelling evidence that the CDW observed in BaFe$_2$Al$_9$ originates from electronic factors.

\section{Computational details}

The first-principles calculations were conducted using the Vienna ab initio simulation package (VASP)~\cite{PhysRevB.47.558, PhysRevB.54.11169}.
The projector augmented wave pseudopotentials recommended by VASP were used~\cite{PhysRevB.50.17953, PhysRevB.59.1758}.
An energy cutoff of 500 eV was used for the plane-wave basis set.
A $\Gamma$-centered $k$-point grid $9 \times9\times13$ was employed for sampling the Brillouin zone.
The convergence criteria for the electronic optimization and structural relaxation were set to 10$^{-7}$ eV and  0.5 meV/$\AA$, respectively.

The phonon dispersions were obtained using a $2\times2\times3$ supercell \textcolor{black}{and a $k$-point grid $4 \times4\times3$},
following the frozen phonon method as implemented in the phonopy code~\cite{Togo2015FirstPP}.
\color{black}
Variations of the supercell size and $k$-point density resulted in slight modifications in the phonon dispersions (see Supplementary Material Fig.~S1~\cite{SM}).
The phonon dispersions were also examined using the density functional perturbation theory~\cite{RevModPhys.73.515},
but only minute differences were observed as compared to those obtained using the frozen phonon method (see Supplementary Material Fig.~S2~\cite{SM}).
The Fermi-Dirac smearing function was adopted, with the width of the smearing
being compatible with the electronic temperature of the system~\cite{PhysRevB.92.245131,PhysRevB.103.035411}.
We checked two widths of smearing, i.e., 1.7 meV and 8.6 meV, which correspond to the electronic temperatures of 20 K and 100 K, respectively.
The results show that varying the widths of smearing only slightly changed the phonon dispersions (see Supplementary Material Fig.~S3~\cite{SM}).
\color{black}

To check electronic correlation effects, we also performed DFT+$U$ calculations based on the Dudarev's scheme~\cite{PhysRevB.57.1505}.
Two exchange-correlation functionals were assessed, namely,
the local density approximation (LDA)~\cite{PhysRevB.23.5048} in the parametrization of Ceperly and Alder~\cite{PhysRevLett.45.566}
and the generalized gradient approximation parameterized by Perdew-Burke-Ernzerhof (PBE)~\cite{PhysRevLett.77.3865}.

\color{black}
The EPC calculations were performed using the EPW code~\cite{PONCE2016116}
as implemented in the QUANTUM ESPRESSO  package~\cite{Giannozzi_2009,Giannozzi_2017}.
Norm-conserving  pseudopotentials~\cite{PhysRevB.88.085117} with an energy cutoff of 55 Ry were used.
The structures were relaxed until the energy and force were smaller than 10$^{-9}$ Ry and 10$^{-7}$ Ry/bohr, respectively.
The Wannier functions were constructed using the Ba-$s$, Al-$sp$, and Fe/Co-$d$ states,
which yield excellent agreement between DFT calculated bands and Wannier interpolated bands (see Supplementary Material Fig.~S4~\cite{SM}).
The  electronic susceptibility and $\bf q$-dependent EPC strength were computed using a 31$\times$31 grid for both $q$- and $k$-points at each plane of a fixed $q_z$.
\color{black}

\section{RESULTS AND DISCUSSION}

\subsection{Crystal structure}

\begin{figure}
\begin{center}
\includegraphics[width=0.49\textwidth, clip]{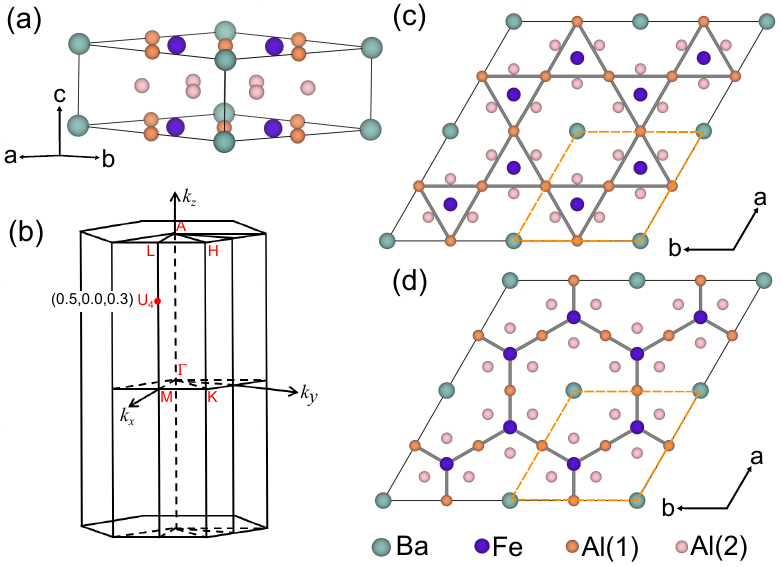}
\end{center}
\caption{(a) Side view of the crystal structure of BaFe$_{2}$Al$_{9}$.
Two nonequivalent crystallographic aluminum sites are highlighted.
(b) The Brillouin zone. The U$_4$ point is located along the L-M path with a fractional coordinate of (0.5, 0.0, 0.3).
(c) Illustration of the hidden kagome lattice formed by Al(1) atoms.
(d) Illustration of the hidden honeycomb lattice formed by Fe atoms.
The yellow dashed lines in (c) and (d) indicate the unit cell.}
\label{Fig1_structure}
\end{figure}

\begin{table}
\caption{The lattice constants and volumes of  BaFe$_{2}$Al$_9$ and BaCo$_{2}$Al$_9$
predicted by LDA and PBE. The experimental data are given for comparison.}
\begin{ruledtabular}
\begin{tabular}{ccccc}
   & & $a$=$b$ (\AA)  &$c$ (\AA) & Volume (\AA$^3$)  \\
\hline
\multirow{3}{0.7cm}{BaFe$_{2}$Al$_9$}
& LDA & 7.87 & 3.87  & 207.9    \\
& PBE  & 8.01 & 3.94  & 219.1   \\
& Expt.~\cite{TURBAN197591}  &  8.04 & 3.89  & 217.8  \\
\hline
\multirow{3}{0.7cm}{BaCo$_{2}$Al$_9$}
& LDA& 7.79 & 3.89  & 205.1   \\
& PBE  & 7.94 & 3.97  & 216.3  \\
& Expt.~\cite{TURBAN197591}  & 7.91 & 3.94  & 213.5   \\
\end{tabular}
\end{ruledtabular}
\label{Table1}
\end{table}

BaFe$_{2}$Al$_{9}$ exhibits a 3D hexagonal structure with a space group of $P6mmm$~\cite{2021_BaFeAl,RYZYNSKA2020121509},
as depicted in Fig.~\ref{Fig1_structure}. Within the crystal structure, two nonequivalent crystallographic Al sites are present,
denoted as Al(1) and Al(2), respectively~\cite{RYZYNSKA2020121509}.
The two nonequivalent crystallographic Al sites have been well resolved
by a recent $^{27}$Al nuclear magnetic resonance (NMR) spectroscopy~\cite{PhysRevB.106.195101}.
On the $ab$ plane, the Al(1) atoms form a kagome lattice [Fig.~\ref{Fig1_structure}(c)],
while the Fe atoms from a honeycomb lattice centered by the Ba atoms [Fig.~\ref{Fig1_structure}(d)].
The hidden kagome and honeycomb substructures in BaFe$_2$Al$_9$
facilitates the emergence of various fascinating electronic and phononic properties,
which will be explored and discussed later.
The plane with Fe, Ba, and Al(1) atoms is separated by the Al(2) atoms.
As compared to the Al(2) atoms, the Al(1) atoms show a stronger bonding
with the Fe atoms within the $ab$ plane, leading to the shorter Fe-Al(1) bonds~\cite{PhysRevB.106.195101}.
BaCo$_{2}$Al$_{9}$ exhibits a structure akin to that of BaFe$_{2}$Al$_{9}$.

The predicted lattice constants and volumes of  BaFe$_{2}$Al$_9$ and BaCo$_{2}$Al$_9$
by LDA and PBE are shown in Table~\ref{Table1}, showing good agreement with the experimental data.
As expected,  the lattice constants are overestimated by PBE, while they are underestimated by LDA.
PBE slightly outperforms LDA in describing the structural properties.

\subsection{Effect of electronic correlation}

\begin{figure}[ht!]
\begin{center}
\includegraphics[width=0.40\textwidth, clip]{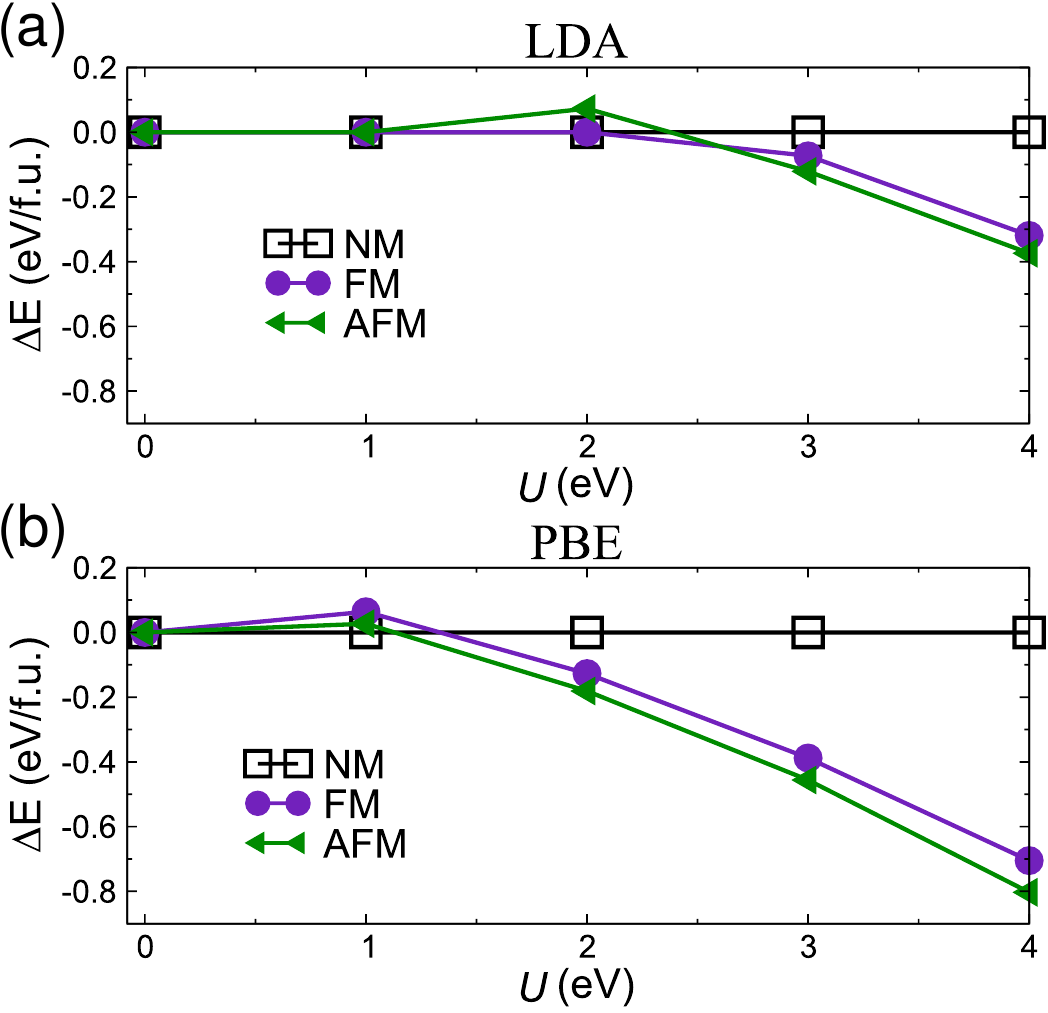}
\end{center}
\caption{The energies of different magnetic configurations of BaFe$_{2}$Al$_{9}$ as a function of $U$
calculated using (a) LDA and (b) PBE functionals. The energy of the respective nonmagnetic phase is aligned to zero.
}
\label{Fig2_E_vs_U}
\end{figure}

We start by identifying the electronic ground states of BaFe$_{2}$Al$_{9}$ and BaCo$_{2}$Al$_{9}$.
Considering the presence of the $3d$ states in both compounds, we examined the electronic correlation effect
by performing DFT+$U$ calculations. Three magnetic configurations [i.e., nonmagnetic (NM), ferromagnetic (FM) and antiferromagnetic (AFM)] were considered.
Fig.~\ref{Fig2_E_vs_U}(a) shows the LDA-predicted energies of the three magnetic configurations of BaFe$_{2}$Al$_{9}$ as a function of $U$.
It is evident that both the AFM and FM configurations reduce to the NM state at $U$ values smaller than 1 eV.
As $U$ increases up to 2 eV, the NM phase remains the most energetically favored one.
As the $U$ value increases further, the magnetically ordered phases become favored,
with the AFM configuration being the most stable one due to the superexchange coupling of the Fe ions mediated by the Al(1) atoms  [Fig.~\ref{Fig1_structure}(d)].
The DFT+$U$ calculations using the PBE functional yield a similar trend as compared to LDA,
but the electronic correlation effect is more pronounced for PBE to stabilize the magnetic order [Fig.~\ref{Fig2_E_vs_U}(b)].
This is a typical behavior for systems with itinerant electrons
where PBE tends to over-stabilize the magnetic ordered phases than LDA~\cite{PhysRevMaterials.4.045001,PhysRevLett.121.207201}.

In contrast to BaFe$_{2}$Al$_{9}$,  BaCo$_{2}$Al$_{9}$ is always nonmagnetic
regardless of the variation of $U$ and the choice of density functionals.
All the configurations with initial magnetic ordered moments reduce to the nonmagnetic state after electronic optimizations.
This is expected, since the Co-$3d$ orbitals are fully filled.
Experimentally, no evidence for the presence of magnetic order were revealed
from magnetic susceptibility, neutron diffraction, and M\"{o}ssbauer results~\cite{2021_BaFeAl}.
Recent $^{27}$Al NMR analyses showed that the magnetic ordering is unlikely
to drive the CDW phase transition in BaFe$_{2}$Al$_{9}$~\cite{PhysRevB.106.195101}.
In addition, we found that applying $U$ worsens the predicted lattice parameters as compared to experiment.
Due to these considerations, our discussions will be limited to the nonmagnetic phase neglecting $U$.
Under this condition, both the PBE and LDA functionals yield similar results.
As a result, only the PBE results will be presented in the subsequent analysis.

\subsection{Phonon dispersions}

\begin{figure}
\begin{center}
\includegraphics[width=0.45\textwidth, clip]{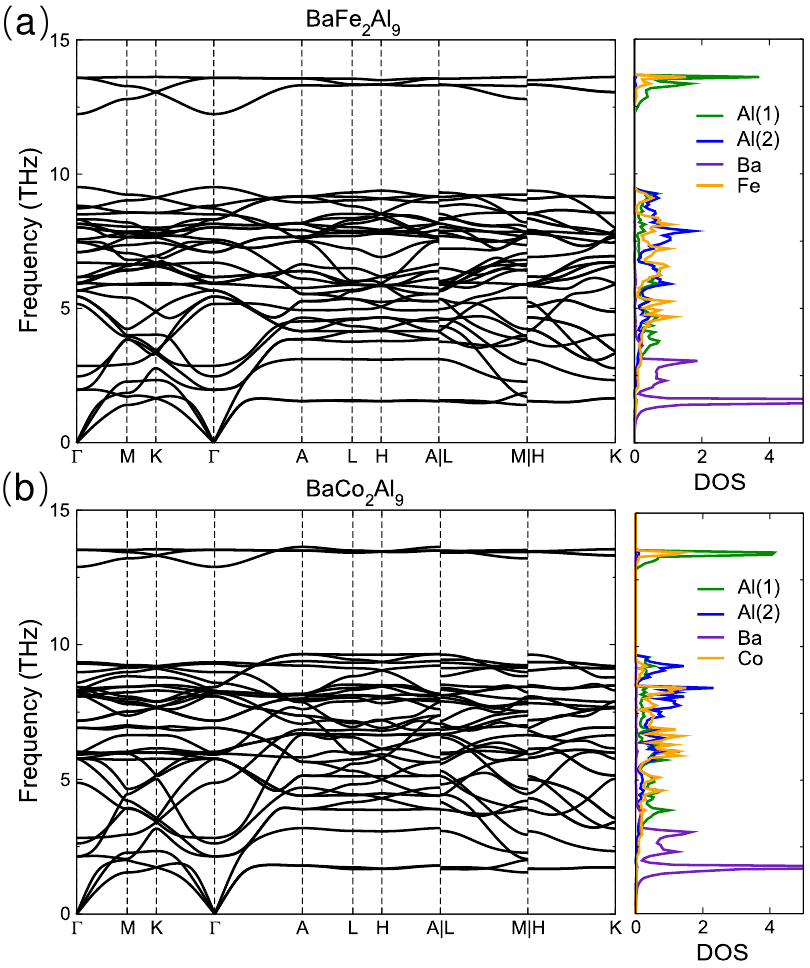}
\end{center}
\caption{Phonon dispersions and projected phononic density of states (in states/THz) of (a) BaFe$_{2}$Al$_{9}$ and (b) BaCo$_{2}$Al$_{9}$
\textcolor{black}{calculated using a width of Fermi smearing of 1.7 meV ($\sim$20 K)}.}
\label{Fig3_phonon_U0}
\end{figure}

The high-temperature pristine phases for a CDW transition often exhibit structural instability,
which is manifested by the presence of soft acoustic phonon
modes~\cite{PhysRevB.94.045131,PhysRevB.98.045114,PhysRevB.101.235405,PhysRevB.92.245131,PhysRevLett.127.046401,PhysRevLett.130.266402}.
By contrast, we found that the high-temperature phases of both BaFe$_{2}$Al$_{9}$ and BaCo$_{2}$Al$_{9}$ are dynamically stable,
which are evidenced by calculated phonon dispersions in Fig.~\ref{Fig3_phonon_U0}.
Besides, both compounds exhibit similar phonon dispersions.
\textcolor{black}{We note that the absence of soft phonon models
is robust against the variation of the $k$-point densities, the widths of Fermi smearing
as well as supercell sizes (see Supplementary Material Fig. S1 and Fig. S3).}
The primary contributors to the acoustic phonons are the vibrations of the heavy Ba atoms.
These phonon bands are rather flat, leading to high phonon density of states (DOSs).
Another flat band around 14 THz and two Dirac bands with linear crossing at $K$ are observed for both compounds,
reminiscent of the typical electronic band structure of a kagome lattice~\cite{Kang2020,Okamoto2022}.
Indeed, these phononic bands come from the Al(1) atoms forming the kagome lattice [Fig.~\ref{Fig1_structure}(c)].
The comparable harmonic phonon dispersions and the absence of soft phonon modes in both compounds
suggest that the CDW transition in BaFe${_2}$Al${_9}$ is not likely to originate from momentum-dependent electron-phonon coupling.
\color{black}
It should be noted that while anharmonic and nonadiabatic effects could lead to the renormalization of phonons~\cite{PhysRevB.82.165111,Alidoosti_2022},
in the present work only the harmonic phonons within the adiabatic regime were considered.
\color{black}

\subsection{Band structure and Fermi surface}

\begin{figure}
\begin{center}
\includegraphics[width=0.49\textwidth, clip]{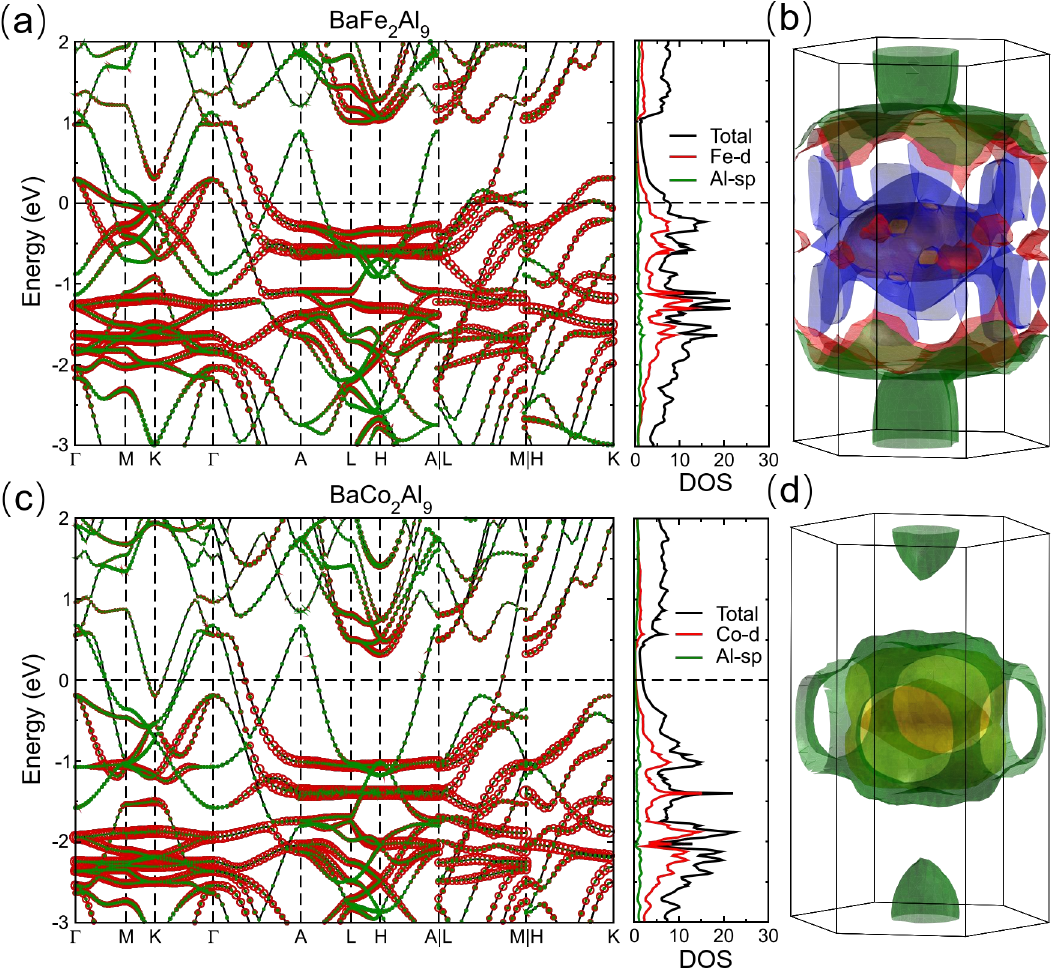}
\end{center}
\caption{Electronic fat band structures and projected density of states (in states/eV) as well as Fermi surfaces
of BaFe$_{2}$Al$_{9}$ [(a)-(b)] and BaCo$_{2}$Al$_{9}$ [(c)-(d)].
The colors in the Fermi surfaces denote the different sheets.
}
\label{Fig4_electronic_U0}
\end{figure}

Figure~\ref{Fig4_electronic_U0} presents the calculated electronic structures and  Fermi surfaces of BaFe$_{2}$Al$_{9}$ and BaCo$_{2}$Al$_{9}$.
The fat band analysis demonstrates their distinct band structures.
The striking difference between the two compounds lies in that the Co-$3d$ orbitals are fully filled,
whereas the Fe-$3d$ orbitals are partially filled.
This results in a significant downward shift ($\sim$0.8 eV) of $d$ bands in BaCo$_{2}$Al$_{9}$.
Both compounds exhibit nearly dispersionless $d$ bands at the $k_z$=0.0 ($\Gamma$-$K$-$M$-$\Gamma$) and $k_z$=0.5 ($A$-$L$-$H$-$A$) planes.
The presence of flat bands and Dirac-like dispersions has been recently identified in BaCo$_{2}$Al$_{9}$
through a combination of angle-resolved photoelectron spectroscopy measurements and density functional theory calculations~\cite{PhysRevB.108.075148}.
Similar to the phononic bands, the electronic flat band (at about $-$1.4 eV for BaFe$_{2}$Al$_{9}$ and $-$2.1 eV for BaCo$_{2}$Al$_{9}$)
and two Dirac bands associated with the hidden Al(1) kagome lattice are also observed around the $K$ point.

In the case of BaCo$_{2}$Al$_{9}$, only three bands intersect the Fermi level, whereas for BaFe$_{2}$Al$_{9}$, multiple bands cross the Fermi level.
This leads to a more complex Fermi surface in BaFe$_{2}$Al$_{9}$ with multiple sheets [compare Fig.~\ref{Fig4_electronic_U0}(b) to Fig.~\ref{Fig4_electronic_U0}(d)].
As compared to BaCo$_{2}$Al$_{9}$, larger electronic DOSs with dominant Fe-$3d$ states at the Fermi level are evident in BaFe$_{2}$Al$_{9}$,
which are mainly contributed by the presence of van Hove singularity (saddle point) near the $M$ point and the point between $L$ and $M$ points.
More detailed presentation on individual Fermi surface sheets are provided in Supplementary Material Fig.~S5~\cite{SM}.

Our results align well with the electronic structure calculations conducted by Meier \emph{et al.}~\cite{2021_BaFeAl},
who proposed that the partially filled Fe-$3d$ bands are responsible for the CDW in BaFe$_{2}$Al$_{9}$~\cite{2021_BaFeAl}.
They hypothesized that the CDW transition would cause certain partially-filled Fe-$3d$ bands to become gapped at the Fermi level,
and as a result, the density of states at the Fermi level would decrease, leading to a sudden rise in resistance upon cooling~\cite{2021_BaFeAl}.
This model appears to be sound and is supported by the NMR evidence
showcasing Fermi surface reconstruction in BaFe$_{2}$Al$_{9}$ associated with the phase transition~\cite{PhysRevB.106.195101}.
All these points support the scenario of the Fermi surface nesting as the underlying cause for the CDW in BaFe$_{2}$Al$_{9}$.

\subsection{Electronic susceptibility}

\begin{figure*}
\begin{center}
\includegraphics[width=0.98\textwidth, clip]{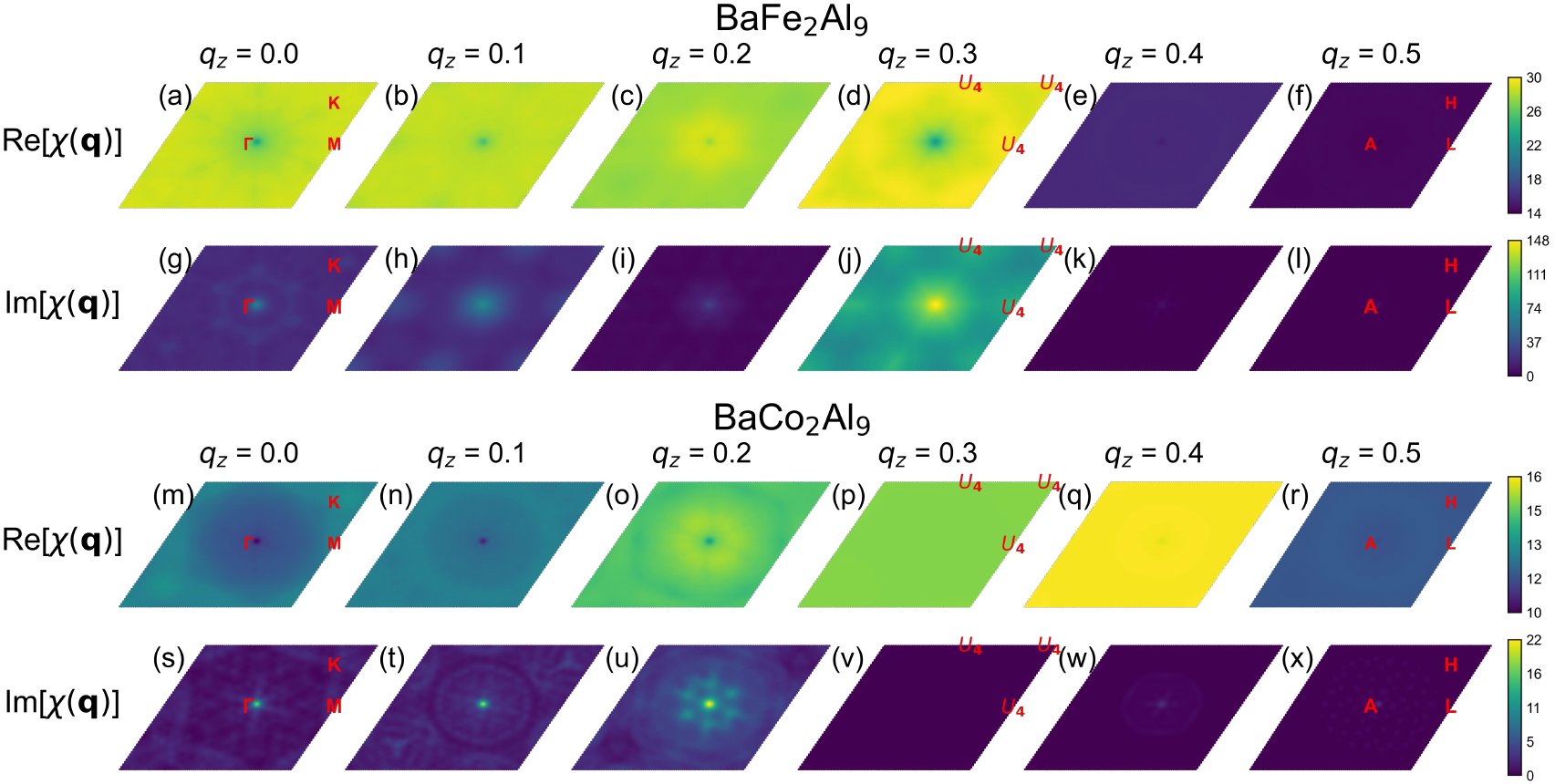}
\end{center}
\caption{The real and imaginary parts of the electronic susceptibilities of
BaFe$_{2}$Al$_{9}$ and BaCo$_{2}$Al$_{9}$ at different $q_z$ planes.
The high-symmetry points and the $U_4$ point are indicated [see Fig.~\ref{Fig1_structure}(b) for the Brillouin zone].
\textcolor{black}{Note the different scales of color bars.}
}
\label{Fig5_FSN_U0}
\end{figure*}

To confirm the Fermi surface nesting mechanism, we computed the  electronic susceptibility of both compounds.
The electronic susceptibility consists of both real and imaginary components, which  represent the response of the system to static and dynamic perturbations, respectively.
The real part of the electronic susceptibility is defined as~\cite{Zong2021}
\begin{equation}\label{eq:re}
{\rm Re[}\chi (\mathbf{q}){\rm ]} = \sum_\mathbf{k} \frac{f(\varepsilon_\mathbf{k}) - f(\varepsilon_{\mathbf{k}+\mathbf{q}})}{\varepsilon_\mathbf{k} - \varepsilon_{\mathbf{k}+\mathbf{q}}},
\end{equation}
where $f(\varepsilon_\mathbf{k})$ is the Fermi-Dirac distribution function at the energy $\varepsilon_\mathbf{k}$.
The imaginary part of the electronic susceptibility (also called ``nesting function") can be calculated by~\cite{Zong2021}
\begin{equation}\label{eq:im}
{\rm Im[}\chi (\mathbf{q}){\rm ]}= \sum_\mathbf{k} \delta(\varepsilon_\mathbf{k} - \varepsilon_{\rm F})\delta(\varepsilon_{\mathbf{k}+\mathbf{q}} - \varepsilon_{\rm F}),
\end{equation}
where $\delta$ is the delta function and $\varepsilon_{\rm F}$ is the Fermi energy.
 ${\rm Re[}\chi (\mathbf{q}){\rm ]}$ captures electronic information both above and below the Fermi surface
and its peak indicates the instability of the electronic system,
whereas ${\rm Im[}\chi (\mathbf{q}){\rm ]}$ directly reflects the Fermi surface topology~\cite{PhysRevB.77.165135,PhysRevResearch.5.013218}.
It is important to note that the CDW instability associated with  the Fermi surface nesting
necessitates  the simultaneous divergence of both components of the electronic susceptibility at the CDW wave vector~\cite{PhysRevB.77.165135}.
Under this prefect nesting condition, the system is highly responsive to both static and dynamic perturbations,
resulting in a reorganization of the electronic states.

Figure~\ref{Fig5_FSN_U0} illustrates the comparison of the calculated electronic susceptibilities between BaFe${_2}$Al${_9}$ and BaCo${_2}$Al${_9}$.
Both the real and imaginary components of the electronic susceptibility at selected $q_z$ planes are displayed.
\color{black}
One can see that BaFe${_2}$Al${_9}$ shows generally higher electronic susceptibilities for both the real and imaginary components as compared to BaCo${_2}$Al${_9}$.
For BaFe${_2}$Al${_9}$, the $q_z$=0.3 plane demonstrates the highest electronic susceptibilities among the considered $q_z$ planes.
\color{black}
In particular, both ${\rm Re[}\chi (\mathbf{q}){\rm ]}$ and ${\rm Im[}\chi (\mathbf{q}){\rm ]}$ reach their maximum at a wave vector $U_4$=(0.5, 0, 0.3) in BaFe${_2}$Al${_9}$.
\textcolor{black}{A clearer illustration through adjusting the color bar scales is presented  in Supplementary Material Fig.~S6~\cite{SM}.}
However, this is not observed in BaCo${_2}$Al${_9}$.
Note that the wave vectors (0.5, 0, $q_z$), (0, 0.5, $q_z$), and (0.5, 0.5, $q_z$) are related by symmetry.
It is worth emphasizing that the wave vector (0.5, 0, 0.3) is the only one where the simultaneous divergence
of both ${\rm Re[}\chi (\mathbf{q}){\rm ]}$ and ${\rm Im[}\chi (\mathbf{q}){\rm ]}$ has been observed in BaFe${_2}$Al${_9}$.
This indicates the strong electronic instability of the high-temperature BaFe${_2}$Al${_9}$ at $U_4$=(0.5, 0, 0.3) towards the CDW transition.
Our calculated electronic susceptibilities are in excellent agreement with the experimental single-crystal X-ray diffraction analysis
showing the incommensurate CDW modulation of BaFe${_2}$Al${_9}$ at a wave vector (0.5, 0, $q_z$) with $q_z$=0.302~\cite{2021_BaFeAl},
and demonstrate that the CDW observed in BaFe$_2$Al$_9$ is most likely of electronic origin.
We note in passing that the electronic instability of BaFe${_2}$Al${_9}$ would be much weakened
if an antiferromagnetic order was considered by applying a $U$=3 eV (see Supplementary Material Fig.~S7~\cite{SM}).
This observation strengthens the notion that the magnetic order is not the underlying cause of the CDW in BaFe$_2$Al$_9$,
consistent with the experimental findings~\cite{2021_BaFeAl,PhysRevB.106.195101}.

\color{black}
\subsection{Electron-phonon coupling}

\begin{figure*}
\begin{center}
\includegraphics[width=0.96\textwidth, clip]{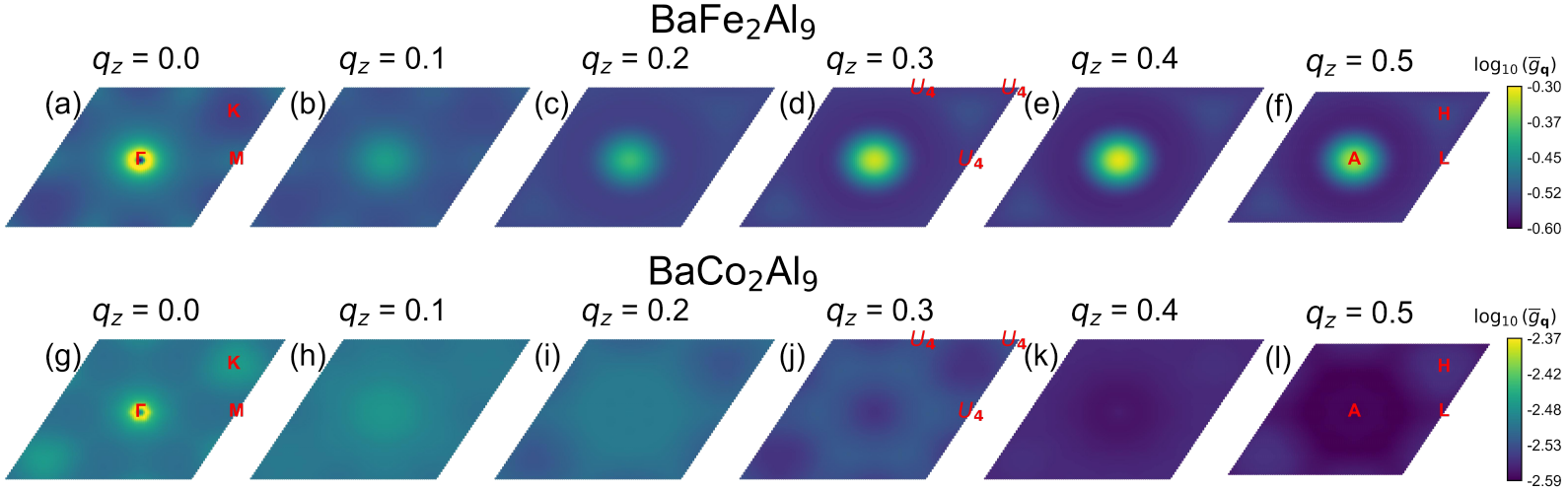}
\end{center}
\color{black}
\caption{The $\bf q$-dependent EPC strength at different $q_z$ planes for BaFe$_2$Al$_9$ (first row) and BaCo$_2$Al$_9$  (second row).
The high-symmetry points and the $U_4$ point are indicated [see Fig.~\ref{Fig1_structure}(b) for the Brillouin zone].
Note that the color bar uses a logarithmic scale.
}
\color{black}
\label{Fig6_EPC}
\end{figure*}

In the previous section, we have shown that the Fermi surface nesting plays an important role in the formation of the CDW in BaFe$_2$Al$_9$.
Considering that the phonons can be renormalized by the EPC~\cite{PhysRevB.80.241108,2015Classification,PhysRevResearch.5.013218},
we computed the $\bf q$-dependent EPC strength by~\cite{PhysRevResearch.5.013218}
\begin{equation}\label{eq:EPC}
\bar{g}_\mathbf{q} = \sum_\mathbf{k} | g_{\mathbf{k},\mathbf{k}+\mathbf{q}}|^2,
\end{equation}
where $g_{\mathbf{k},\mathbf{k}+\mathbf{q}}$ is the EPC matrix element,
which couples the electronic states
at momentums of $\mathbf{k}$ and $\mathbf{k}+\mathbf{q}$ with a phonon with a momentum of $\mathbf{q}$.

Figure~\ref{Fig6_EPC} presents the calculated $\bf q$-dependent EPC strength $\bar{g}_\mathbf{q}$ at selected $q_z$ planes for BaFe$_2$Al$_9$ and BaCo$_2$Al$_9$.
One can observe that the EPC strength in BaFe$_2$Al$_9$
generally exceeds that in BaCo$_2$Al$_9$ by nearly two orders of magnitude (note the logarithmic scale color bars).
Given the very similar phonon dispersions of the two compounds (see Fig.~\ref{Fig3_phonon_U0}),
the enhanced EPC strength observed in BaFe$_2$Al$_9$
must originate from its distinct electronic bands crossing the Fermi energy
that lead to multiple sheets of the Fermi surface and high density of states at the Fermi level (see Fig.~\ref{Fig4_electronic_U0}).
\textcolor{black}{Although the CDW phase would be expected in BaFe$_2$Al$_9$ from the computed electronic susceptibilities,
the EPC strength in BaFe$_2$Al$_9$ is nearly negligible and suggests no indication of the CDW.
This is manifested by the absence of a peak at the Fermi surface nesting wave vector $U_4$=(0.5, 0, 0.3),
at which the CDW occurs in experiment~\cite{2021_BaFeAl}.}
These EPC calculations indicate that as compared to the Fermi surface nesting,
the $\bf q$-dependent EPC has a less decisive impact on inducing the CDW in BaFe$_2$Al$_9$.

\color{black}

\subsection{The CDW modulated structure}

Here, we would like to briefly discuss the CDW modulated structure of BaFe$_2$Al$_9$ at low temperature.
In Ref.~\cite{2021_BaFeAl} a superlattice structure model consisting of 2$\times$1$\times$10 conventional unit cells was proposed.
In this model, the Ba toms along the [001] direction develop a sine-modulated horizontal displacement with a period of 10/3$c$ ($c$ being the lattice constant of unit cell)
and the Fe atoms exhibit modulated displacements along the [001] direction, resulting in varying Fe-Fe distances~\cite{2021_BaFeAl}.
Following Ref.~\cite{2021_BaFeAl} we constructed such sine-modulated superlattice using the ISODISTORT tool~\cite{ISODISTORT} according to the $U_4$ displacement modes
and fully relaxed the structure.
It turned out that the resulting superlattice closely resembles the high-temperature phase.
This is evident from the distortion amplitude,
which was determined to be nearly negligible ($\sim$0.002 $\AA$) compared to the high-temperature phase using the AMPLIMODES tool~\cite{AMPLIMODES}
(see Supplementary Material Fig.~S8~\cite{SM}).
This implies that the proposed sine-modulated model does not represent the true low-temperature structure.
Indeed, it was pointed out that the sine-modulated model derived from the single-crystal X-ray diffraction refinement
exhibits an inconsistency with the M\"{o}ssbauer result that instead better supports a two Fe-site model~\cite{2021_BaFeAl}.
Given the incommensurate \textcolor{black}{and three-dimensional characteristics of the CDW}
and the possible presence of higher-order harmonics to the CDW modulation in BaFe${_2}$Al${_9}$~\cite{2021_BaFeAl},
identifying its low-temperature CDW structure would be fairly challenging \textcolor{black}{both in experiment and theory}.

\section{Conclusions}

In conclusion, we have conducted a comparative and systematic first-principles study
of the electronic and dynamical properties \textcolor{black}{as well as electron-phonon interactions} of the two
isostructural intermetallics, BaFe${_2}$Al${_9}$ and BaCo${_2}$Al${_9}$, to unravel the
underlying origin of the CDW appearing only in BaFe${_2}$Al${_9}$.
The similar phonon dispersions, absence of soft phonon modes, \textcolor{black}{and weak EPC at the CDW wave vector}
in the high-temperature phase of both compounds indicate that
\textcolor{black}{the $\bf q$-dependent EPC does not play a decisive role in the formation of CDW in BaFe${_2}$Al${_9}$}.
In contrast to BaCo${_2}$Al${_9}$ where the Co-$3d$ states are fully filled,
the Fe-$3d$ states are partially filled in BaFe$_2$Al$_9$,
This results in multiple bands crossing the Fermi level, higher density of states at the Fermi level,
and a more complex Fermi surface in BaFe$_2$Al$_9$.
Through electronic susceptibility calculations,
we have identified that only BaFe$_2$Al$_9$ exhibits a simultaneous divergence of
both the real and imaginary parts at the CDW wave vector ${\bf q}$=(0.5, 0, 0.3), which is absent in BaCo$_2$Al$_9$.
Our findings align remarkably well with experimental observations,
establishing that the observed CDW in BaFe$_2$Al$_9$
is primarily driven by the electronic instability associated with the Fermi surface nesting.

\section*{Acknowledgements}
P.L. thanks Yanxu Wang for useful discussions.
This work is supported by
the National Key R\&D Program of China 2021YFB3501503,
the National Natural Science Foundation of China  (Grants No. 52422112, No. 52188101, and No. 52201030),
the Liaoning Province Science and Technology Planning Project (2023021207-JH26/103 and RC230958),
and the Special Projects of the Central Government in Guidance of Local Science and Technology Development (2024010859-JH6/1006).
Part of the numerical calculations in this study were carried out on the ORISE Supercomputer (Grant No. DFZX202319).

\bibliography{Reference} 

\end{document}


\title{Supplemental Material to \\
		``Origin of the charge density wave state in BaFe$_2$Al$_9$"}

\author{Yuping Li}
\thanks{These authors contribute equally to this work.}
\affiliation{Key Laboratory for Anisotropy and Texture of Materials (Ministry of Education),
		School of Materials Science and Engineering, Northeastern University, Shenyang 110819, China}
\affiliation{%
Shenyang National Laboratory for Materials Science, Institute of Metal Research, Chinese Academy of Sciences, 110016 Shenyang, China
}%

\author{Mingfeng Liu}
\thanks{These authors contribute equally to this work.}
\affiliation{%
Shenyang National Laboratory for Materials Science, Institute of Metal Research, Chinese Academy of Sciences, 110016 Shenyang, China
}%

\author{Jiangxu Li}
\email{jxli15s@imr.ac.cn}
\affiliation{%
Shenyang National Laboratory for Materials Science, Institute of Metal Research, Chinese Academy of Sciences, 110016 Shenyang, China
}%

\author{Jiantao Wang}
\affiliation{%
Shenyang National Laboratory for Materials Science, Institute of Metal Research, Chinese Academy of Sciences, 110016 Shenyang, China
}%
\affiliation{%
School of Materials Science and Engineering, University of Science and Technology of China, 110016 Shenyang, China
}%

\author{Junwen Lai}
\affiliation{%
Shenyang National Laboratory for Materials Science, Institute of Metal Research, Chinese Academy of Sciences, 110016 Shenyang, China
}%
\affiliation{%
School of Materials Science and Engineering, University of Science and Technology of China, 110016 Shenyang, China
}%

\author{Dongchang He}
\affiliation{%
Shenyang National Laboratory for Materials Science, Institute of Metal Research, Chinese Academy of Sciences, 110016 Shenyang, China
}%
\affiliation{%
School of Materials Science and Engineering, University of Science and Technology of China, 110016 Shenyang, China
}%

\author{Ruizhi Qiu}
\affiliation{%
Institute of Materials, China Academy of Engineering Physics, Mianyang 621907, China
}%

\author{Yan Sun}
\affiliation{%
Shenyang National Laboratory for Materials Science, Institute of Metal Research, Chinese Academy of Sciences, 110016 Shenyang, China
}%

\author{Xing-Qiu Chen}%
\email{xingqiu.chen@imr.ac.cn}
\affiliation{%
Shenyang National Laboratory for Materials Science, Institute of Metal Research, Chinese Academy of Sciences, 110016 Shenyang, China
}%

\author{Peitao Liu}%
\email{ptliu@imr.ac.cn}
\affiliation{%
Shenyang National Laboratory for Materials Science, Institute of Metal Research, Chinese Academy of Sciences, 110016 Shenyang, China
}%

\maketitle

	\begin{figure}
		\begin{center}
			\includegraphics[width=0.9\textwidth, clip]{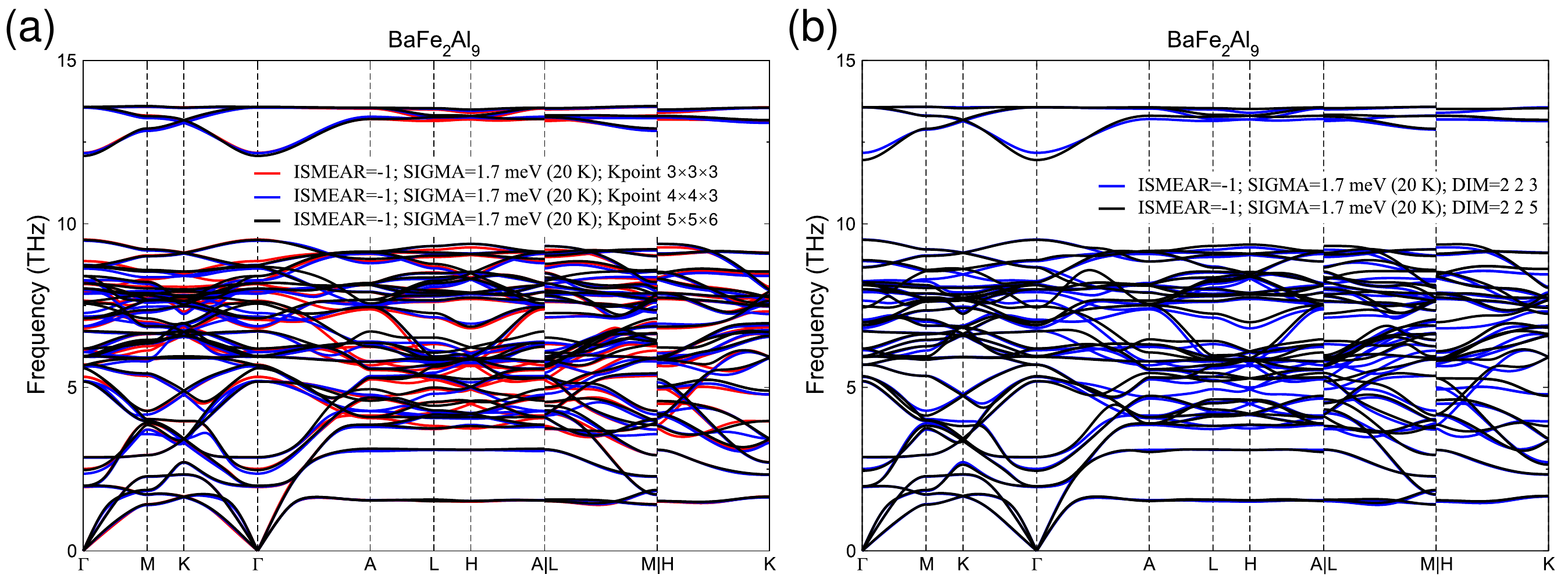}
		\end{center}
		\caption{Dependence of the calculated phonon dispersions of BaFe$_{2}$Al$_{9}$
on (a) the $k$-point density for a fixed $2\times2\times3$ supercell
and (b) the supercell size for a fixed $4\times4\times3$ $k$-point grid.
}
		\label{FigS1}
\end{figure}

	\begin{figure}
		\begin{center}
			\includegraphics[width=0.9\textwidth, clip]{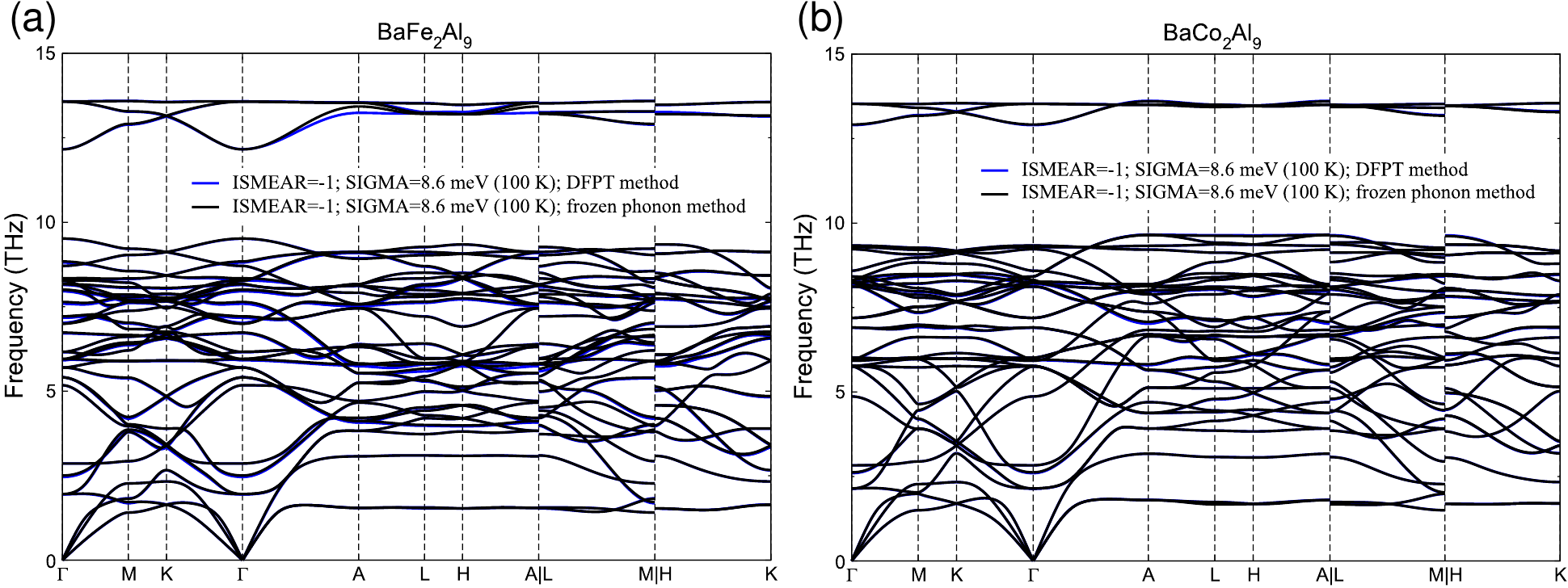}
		\end{center}
		\caption{Calculated phonon dispersions of (a) BaFe$_{2}$Al$_{9}$ and (b) BaCo$_{2}$Al$_{9}$
 using the DFPT method (blue lines) or the frozen phonon method  (black lines).
}
		\label{FigS2}
\end{figure}

	\begin{figure}
		\begin{center}
			\includegraphics[width=0.9\textwidth, clip]{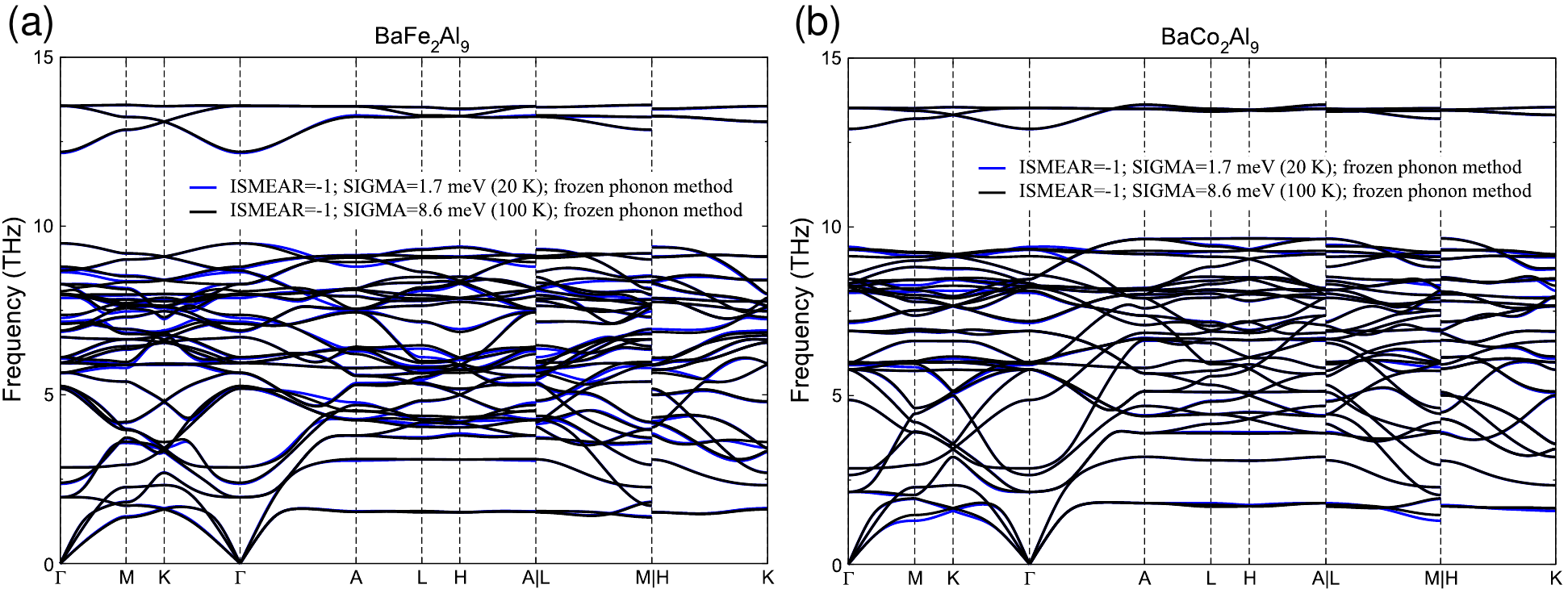}
		\end{center}
		\caption{Calculated phonon dispersions of (a) BaFe$_{2}$Al$_{9}$ and (b) BaCo$_{2}$Al$_{9}$
 using widths of Fermi smearing of 1.7 meV  (blue lines) or 8.6 meV  (black lines).
}
		\label{FigS3}
\end{figure}

	\begin{figure}
		\begin{center}
			\includegraphics[width=0.9\textwidth, clip]{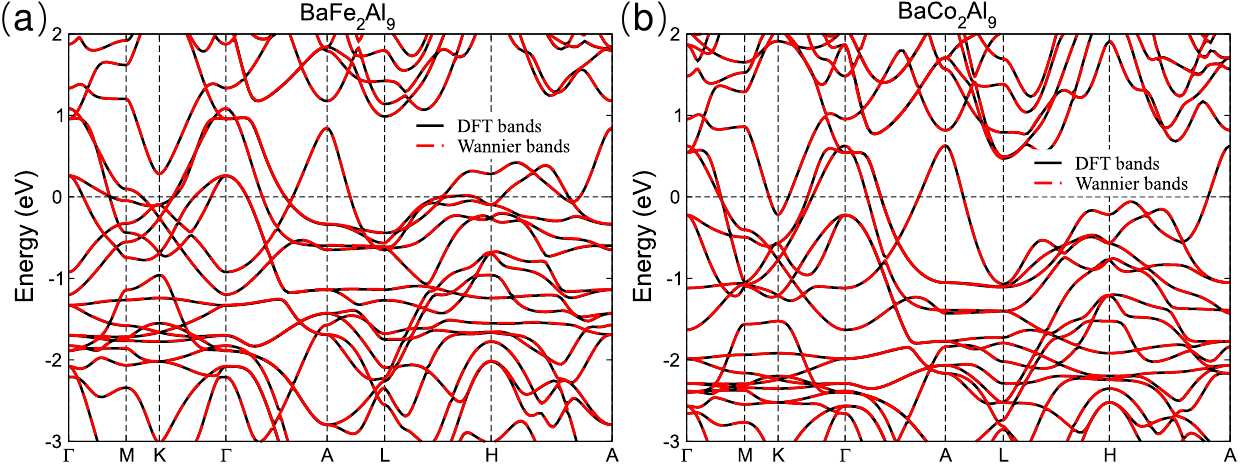}
		\end{center}
		\caption{Comparison of DFT-calculated and Wannier-interpolated electronic bands of  (a) BaFe$_{2}$Al$_{9}$ and (b) BaCo$_{2}$Al$_{9}$.}
		\label{FigS4}
\end{figure}

	\begin{figure}
		\begin{center}
			\includegraphics[width=0.95\textwidth, clip]{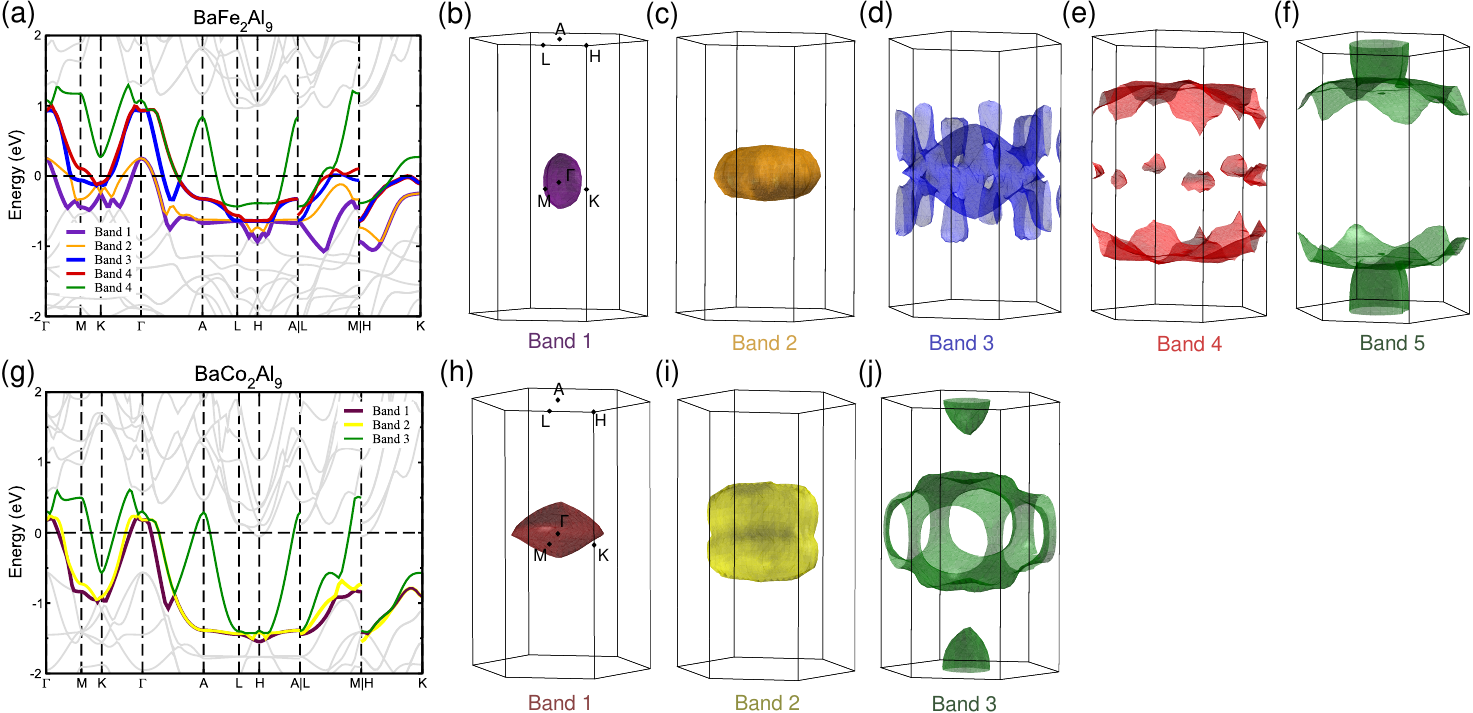}
		\end{center}
		\caption{The electronic band structures and the individual Fermi surface sheets of BaFe$_{2}$Al$_{9}$ (upper panels) and BaCo$_{2}$Al$_{9}$ (bottom panels),
}
		\label{FigS5}
\end{figure}

	\begin{figure}
		\begin{center}
			\includegraphics[width=0.98\textwidth, clip]{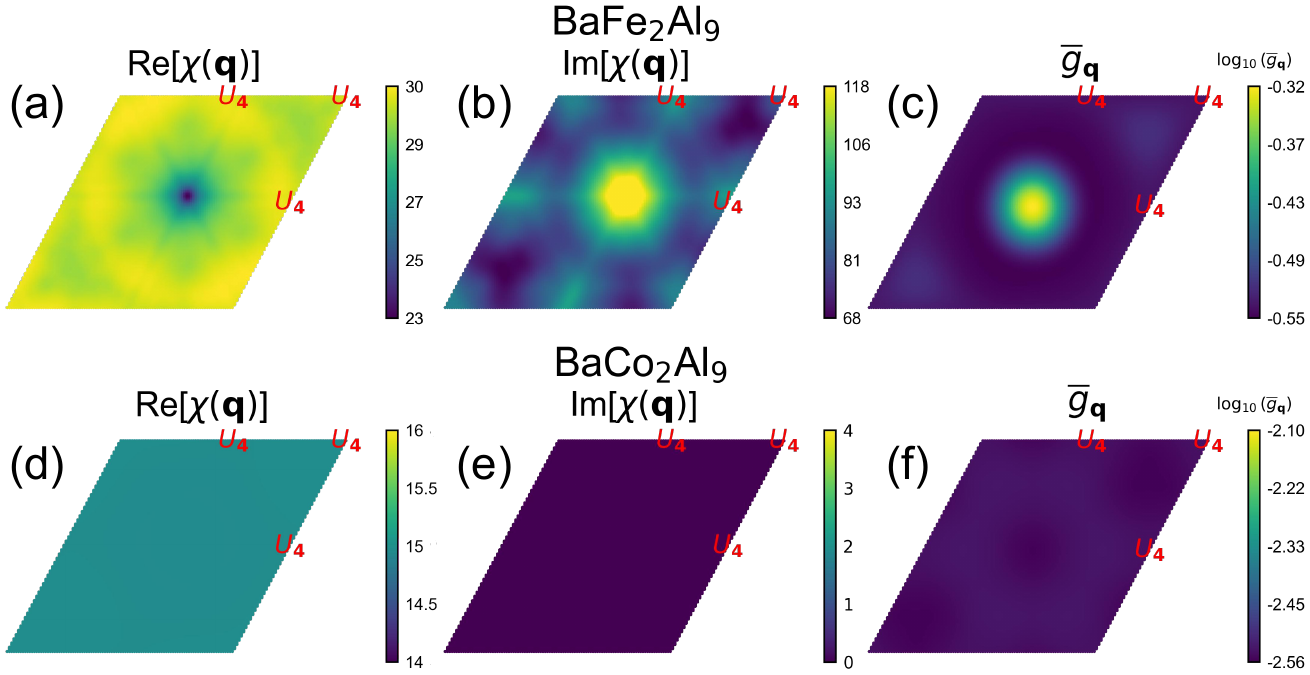}
		\end{center}
		\caption{The real (${\rm Re[}\chi (\mathbf{q}){\rm ]}$) and imaginary (${\rm Im[}\chi (\mathbf{q}){\rm ]}$)
parts of the electronic susceptibilities and the $\bf q$-dependent electron-phonon coupling strength ($\bar{g}_\mathbf{q}$)
of BaFe$_{2}$Al$_{9}$ and BaCo$_{2}$Al$_{9}$ at the $q_z$=0.3 plane.
		}
		\label{FigS6}
	\end{figure}

	\begin{figure}
		\begin{center}
			\includegraphics[width=0.98\textwidth, clip]{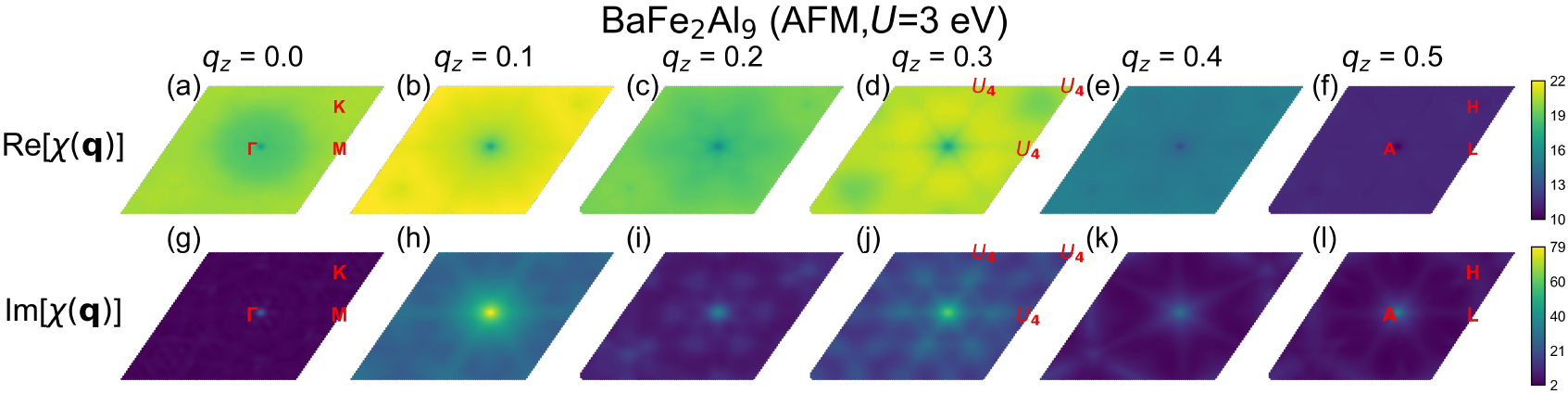}
		\end{center}
		\caption{The real and imaginary parts of the electronic susceptibilities of
BaFe$_{2}$Al$_{9}$ with an antiferromagnetic (AFM) order obtained using $U$=3 eV.
		}
		\label{FigS7}
	\end{figure}

	\begin{figure}
		\begin{center}
			\includegraphics[width=0.70\textwidth, clip]{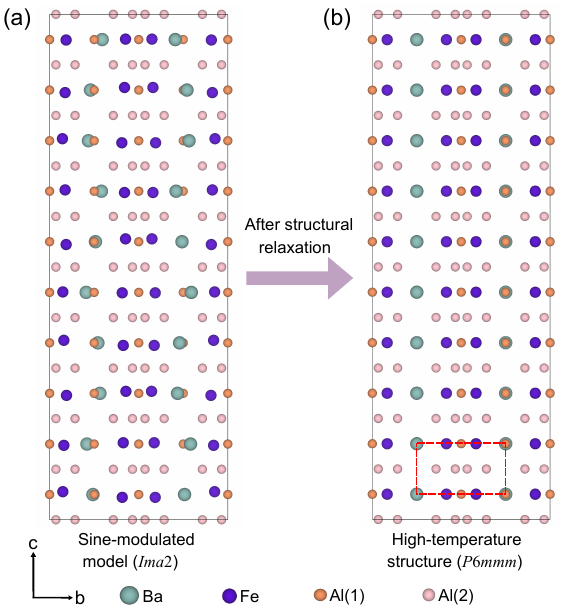}
		\end{center}
		\caption{(a) The sine-modulated superlattice model consisting of 2$\times$1$\times$10 conventional unit cells (space group  $Ima2$),
 which was constructed according to the $U_4$ displacement modes.
(b) The structure after full relaxation, which closely resembles the high-temperature phase (space group $P6mmm$).
The red dashed rectangle denotes the conventional unit cell of  the high-temperature phase.
		}
		\label{FigS8}
	\end{figure}
